\documentclass[12pt]{article}
\usepackage{graphicx}
\usepackage{amssymb, amsmath, amsfonts}

\def\drawbox#1#2{\hrule height#2pt
        \hbox{\vrule width#2pt height#1pt \kern#1pt
              \vrule width#2pt}
              \hrule height#2pt}

\def\Asym#1#2{\vcenter{\vbox{\drawbox{#1}{#2}
              \kern-#2pt 
              \drawbox{#1}{#2}}}}

\newcommand{\bear}{\begin{eqnarray}}
\newcommand{\eear}{\end{eqnarray}}

\def\beq{\begin{equation}}
\def\eeq{\end{equation}}
\def\bal{\begin{align}}
\def\eal{\end{align}}

\begin{document}
\title{
  (Near) Conformal Technicolor: \\~\\ What is really new? }
  
  \author{
  Francesco Sannino    \\~\\
  {\em  HEP Center, University of Southern Denmark.} \\{\em Campusvej 55, DK-5230, Odense M., Denmark}}
 
\maketitle

\baselineskip=14pt

\begin{abstract}
 The knowledge of the phase diagram of strongly coupled theories as function of the number of colors, flavors and matter representation plays a fundamental role when constructing viable extensions of the standard model (SM) featuring dynamical electroweak symmetry breaking. Here I summarize the state-of-the-art of the phase diagram for $SU(N)$ gauge theories with fermionic matter transforming according to arbitrary representations of the underlying gauge group. I critically report on the latest results from first principle lattice simulations and then review the principal models of (near) conformal technicolor such as (Next) Minimal Walking Technicolor (MWT) and Partially Gauged Technicolor (PGT).  I finally show that the incarnation of the conformal technicolor model is nothing but the simplest PGT model.  \end{abstract}
\newpage
\baselineskip=18pt
\newpage

\tableofcontents

\newpage

\section{Background}

Models of electroweak symmetry breaking via new strongly interacting
theories of technicolor type \cite{Weinberg:1979bn,Susskind:1978ms}
are gaining momentum. The most updated review on the subject has just appeared \cite{Sannino:2008ha} while earlier ones are \cite{Hill:2002ap,Lane:2002wv}. There is no doubt that the main difficulty in
constructing such extensions
of the SM is the very
limited knowledge about generic strongly interacting theories. This
 has led theorists, in the past, to construct models of technicolor
resembling ordinary quantum chromodynamics  \cite{Weinberg:1979bn,Susskind:1978ms}. Unfortunately the simplest version of this type of models are at odds with electroweak precision measurements. New strongly coupled theories with dynamics very different from the one featured by a scaled up version of QCD are needed as summarized in \cite{Sannino:2008ha}. 

In this mini-review I summarize first the state-of-the-art of the phase diagram \cite{Ryttov:2007cx,Ryttov:2007sr,Dietrich:2006cm} for SU(N)  gauge theories and then present the most recent models of dynamical electroweak symmetry breaking \cite{Sannino:2004qp,Dietrich:2005jn,Dietrich:2005wk,Foadi:2007ue}. 

\section{Phase Diagram}
{}First principle lattice simulations are now capable to  investigate the spectrum and the dynamics of various four dimensional gauge theories which are of interest in our pursue of a dynamical origin of the stabilization of the Fermi scale \cite{DelDebbio:2008wb,Catterall:2007yx,Appelquist:2007hu}.  It is, however, very useful to provide an analytical study of the dynamics and/or spectrum of a generic nonsupersymmetric gauge theory applying, for example, the proposal of the all-order beta function for nonsupersymmetric gauge theories with fermionic matter \cite{Ryttov:2007cx}. This new method constitutes a true step forward with respect to the very rough method based on the truncated Schwinger-Dyson equation (SD) \cite{Appelquist:1988yc,Cohen:1988sq,Miransky:1996pd} (referred also as the ladder approximation in the literature) or even conjectures such as the Appelquist-Cohen-Schmaltz (ACS) one  \cite{Appelquist:1999hr} which makes use of the counting of the thermal degrees of freedom at high and low temperature. The ACS conjecture is, in fact, unable to constrain the phase diagram for vector-like theories with matter in higher dimensional representations as I  have shown in \cite{Sannino:2005sk}. The ACS conjecture has been tested also for chiral gauge theories \cite{Appelquist:1999vs}. There it was also found that to make definite predictions a stronger requirement is needed \cite{Appelquist:2000qg}.

\subsection{All-order beta function}
Let's start from the proposal of the beta function for nonsupersymmetric $SU(N)$ gauge theories with fermionic matter \cite{Ryttov:2007cx}. It is  written in a form useful for constraining the phase diagram of strongly coupled theories. The form is inspired by the Novikov-Shifman-Vainshtein-Zakharov  (NSVZ) beta function for supersymmetric theories \cite{Novikov:1983uc,Shifman:1986zi} and the renormalization scheme coincides with the NSVZ one. We proposed the following form \cite{Ryttov:2007cx} of the beta function:
\begin{eqnarray}
\beta(g) &=&- \frac{g^3}{(4\pi)^2} \frac{\beta_0 - \frac{2}{3}\, T(r)\,N_f \,
\gamma(g^2)}{1- \frac{g^2}{8\pi^2} C_2(G)\left( 1+ \frac{2\beta_0'}{\beta_0} \right)} \ ,
\end{eqnarray}
with
\begin{eqnarray}
\beta_0 =\frac{11}{3}C_2(G)- \frac{4}{3}T(r)N_f \ , \qquad  \beta_0' = C_2(G) - T(r)N_f  \ .
\end{eqnarray}
We have also defined $\gamma =-{d\ln m}/{d\ln \mu}$ and $m$ the renormalized fermion mass. The generators $T_r^a,\, a=1\ldots N^2-1$ of the gauge group in the
representation $r$ are normalized according to
$\mathrm{Tr}\left[T_r^aT_r^b \right] = T(r) \delta^{ab}$ while the
quadratic Casimir $C_2(r)$ is given by $T_r^aT_r^a = C_2(r)I$. The adjoint
representation is denoted by $G$. 
\subsubsection{Free Electric Phase}
This is the region of $N_f$ for which  $\beta_0$ is negative and asymptotic freedom is lost. The theory behaves like QED and hence it becomes strongly coupled at high energy. $N_f^\mathrm{I}$ is the number of flavors above which the theory is no longer asymptotically free. This corresponds to \mbox{$\beta_0[N_f^\mathrm{I}]{=}0$}. For values of $N_f$ larger than $N_f^\mathrm{I}$ the theory is in a non-Abelian QED theory. We obtain
\begin{eqnarray}
N_f^{\rm{I}} = \frac{11}{4} \frac{C_2(G)}{T(r)} \ .
\end{eqnarray}

\subsubsection{Coulomb Phase}
As we decrease the number of flavors from just below the point where asymptotic freedom is lost one expects a perturbative (in the coupling) zero in the beta function to occur \cite{Banks:1981nn}. From the expression proposed above one finds that at the zero of the beta function, barring zeros in the denominator,  one must have
\begin{eqnarray}
\gamma = \frac{11C_2(G)-4T(r)N_f}{2T(r)N_f} \ .\end{eqnarray}
The dimension of the chiral condensate is $D(\bar{\psi} \psi)=3-\gamma$ which at the IR fixed point value reads
\begin{equation}
D (\bar{\psi} \psi)= \frac{10T(r)N_f - 11C_2(G)}{2T(r)N_f} \ .
\end{equation}
 To avoid negative norm states in a conformal field theory one must have $D\geq 1$ for non-trivial
 spinless operators \cite{Mack:1975je,Flato:1983te,Dobrev:1985qv}.

Hence the critical number of flavors below which the unitarity bound is violated is
\begin{eqnarray}
N_f^{\rm{II}} = \frac{11}{8} \frac{C_2(G)}{T(r)} \ ,
\end{eqnarray}
which corresponds to having set $\gamma=2$. The analysis above is
similar to the one for supersymmetric gauge theories \cite{Seiberg:1994pq}.
 The actual size of the conformal window may be smaller than the one
presented here which is the bound on the size of the window. The reason being that chiral symmetry breaking could be triggered for a value of $\gamma$ lower than two. 

A value of $\gamma$ larger than one, still allowed by unitarity, is a welcomed feature when using this window to construct walking technicolor theories \cite{Yamawaki:1985zg,Holdom:1984sk,Holdom:1981rm,Appelquist:1986an}. It may allow for the physical value of the mass of the top while avoiding a large violation of flavor changing neutral currents which were investigated in  \cite{Evans:2005pu} for the minimal walking model.

\subsection{Conformal Window}
I  now compare and combine analytical predictions for the conformal window with lattice results \cite{Catterall:2007yx,Shamir:2008pb,Appelquist:2007hu,DelDebbio:2008zf}. The first exhaustive perturbative analysis relevant to start a systematic study of gauge theories with fermions in any given representation of the $SU(N)$ on the lattice has just appeared \cite{DelDebbio:2008wb}.

\subsubsection{Two-index symmetric representation}
Two and three colors with two Dirac flavors transforming according to the two index symmetric (2S) representation of the gauge group have been investigated on the lattice respectively in \cite{Catterall:2007yx} and \cite{Shamir:2008pb}.
 For $SU(2)$ the spectrum of the theory  \cite{Catterall:2007yx} has been studied and confronted with the theory with two colors and two Dirac flavors in the fundamental representation. The lattice studies indicate that either the theory is very near an infrared stable fixed point or the fixed point is already reached. These are only preliminary results and more refined  investigations are needed (see the section on the Schr\"odinger's zeros). Nevertheless let's compare them directly with analytical results. According to ladder results we should be below the conformal window but very near conformal \cite{Sannino:2004qp}.  According to the all-order beta function the anomalous dimension of the mass operator, if the IR fixed point is reached, assumes the value:
 \begin{equation}
 \gamma = \frac{3}{4}\ ,   \quad SU(2)~~{\rm model~with~2~ (2S)~Flavors.}  \end{equation} 
The all-order beta function shows that one has not yet reached $\gamma$ equal one and suggests that the $SU(2)$ model is indeed conformal in the infrared if one uses $\gamma =1 $ as an indication of when the conformal window ceases to exist. However, as explained above, the constraint coming from unitarity of the conformal theories allows $\gamma$ to take even larger values, i.e. up to 2, before loosing conformality. 

The situation is very intriguing for the $SU(3)$ theory. Recent lattice results  \cite{Shamir:2008pb} suggest that this theory may already have achieved an IR fixed point. Here, as well, more studies are needed. The ladder approximation predicts, however, this theory to be near conformal (i.e. walking) but further away from conformality then the $SU(2)$ theory.  If the theory were indeed conformal in the infrared, via the all-order beta function, we predict the anomalous dimension of the fermion condensate to assume the following value:  
 \begin{equation}
 \gamma = 1.3 \ , \quad SU(3)~~{\rm model~with~2~ (2S)~Flavors.}  \end{equation} 
The anomalous dimension of the mass operator turns out to be larger than one! This would be quite an important result since large anomalous dimensions are needed when constructing extended technicolor models able to account for the heavy quark masses. In fact the common lore is that the anomalous dimension of the quark operator does not exceed one.  If the $SU(3)$ generates an infrared fixed point then the $SU(2)$ would also generate it since fermions screen even more there.

 \subsubsection{Fundamental representation}
The all-order beta function {\it predicts} that the conformal window cannot be achieved for a number of flavors less then 8.25 in the fundamental representation of $SU(3)$. This is supported by the latest lattice results \cite{Deuzeman:2008sc,Appelquist:2007hu}. If this theory develops an infrared fixed point we predict the anomalous dimension of the quark mass operator to be:
\begin{equation}
 \gamma = \frac{3}{4} \ ,  \quad SU(3)~~{\rm model~with~12~Fundamental~Flavors}
 \end{equation} 
Amusingly the theories with 12 fundamental flavors in $SU(3)$ and 2 adjoint Dirac flavors in $SU(2)$ (adjoint fermions here correspond to the 2S in this case) have the same anomalous dimension if both develop the infrared fixed point. What is extremely interesting to know is if a fixed point is generated for a number of flavors less then eleven but higher than eight since according to the all-order beta function this corresponds to an anomalous dimension larger than one but still smaller than two. 

\begin{figure}[h!]
\begin{center}\resizebox{12cm}{!}{\includegraphics{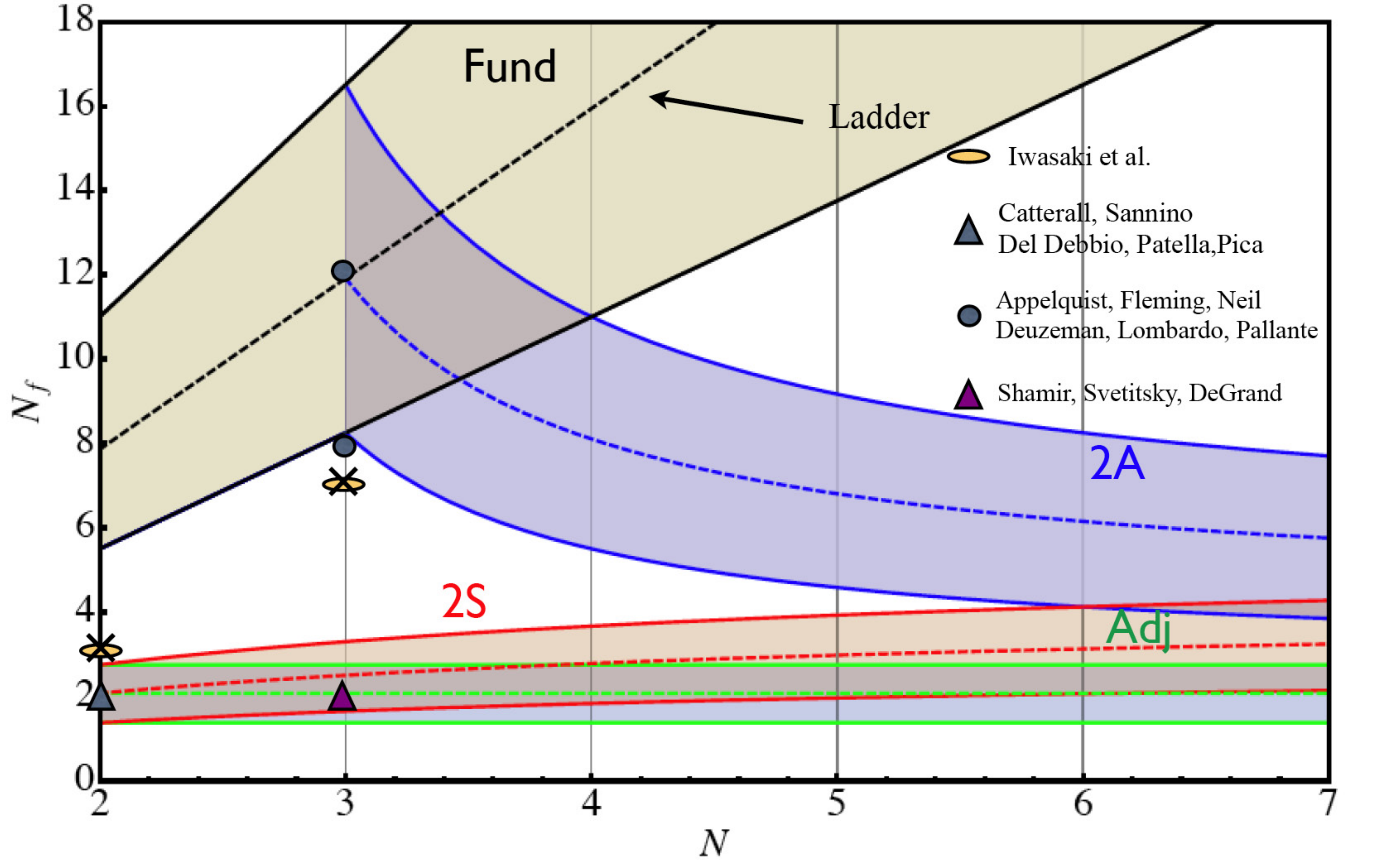}}
\caption{Phase diagram for nonsupersymmetric theories with fermions in various representations with superimposed for which theories the lattice simulations were performed. The dashed lines correspond to the ladder approximation boundary of the conformal window which correspond to gamma about one. While the bound from the all-order beta function is obtained for $\gamma=2$. Note that if we were to use $\gamma=1$ constraint with the all-order beta function the conformal window would be a little larger than the ladder one. Oval and round circles denote early lattice studies \cite{Iwasaki:2003de} with fermions in the fundamental representation. Triangles denote the lattice results for fermions in the two index representation. The cross on the ovals denote that the conclusion of the theories already being conformal  is in disagreement with the theoretical predictions.} \label{LatticeandPD}\end{center}
\end{figure}

The phase diagram is summarized in figure \ref{LatticeandPD}.

\subsection{Schr\"odinger's zeros:  Are they physical?}
The beta function derived on the lattice using the Schr\"odinger functional \cite{Luscher:1992an,Luscher:1993gh,Sint:1993un,Bode:1999sm} has exactly the same limitations of the 't Hooft beta function. {} For example the presence or absence of a zero in these schemes does not demonstrate, per se, the presence or the absence of a physical fixed point.  More information is needed, such as the knowledge of the anomalous dimension of the fermion mass at the fixed point. One can infer the existence of an infrared fixed point by measuring, on the lattice, correlators of gauge-invariant operator  and check if they display power law behaviors.  Scaling of certain relevant quantities which can be measured on the lattice were derived in \cite{Sannino:1999qe} using the implementation of trace anomaly at the effective Lagrangian level.  Since we used scaling arguments the results are as general as the ones in \cite{Luty:2008vs}. Note that differently from the 't Hooft and the Schr\"odinger functional case the all-order beta function presented above predicts the anomalous dimensions at the fixed point. These are physical quantities, i.e. independent from the scheme. 
\section{Better Models of Technicolor }
Having shed light one the phase diagram of strongly coupled theories we are now entitled to investigate possible (near) conformal technicolor models. 

The simplest technicolor model has $N_{T f}$ Dirac fermions in
the fundamental representation of $SU(N)$. These models, when
extended to accommodate the fermion masses through the extended technicolor interactions,
suffer from large flavor changing neutral currents. This problem is
alleviated if the number of flavors is sufficiently large
such that the theory is (almost) conformal. This is estimated to happen, for fermion in the fundamental representation, 
for $N_{T f} \sim 4 N$ \cite{Yamawaki:1985zg}. This, in turn, implies a large
contribution to the oblique parameter $S$ \cite{Hong:2004td} when all of the flavor symmetries are gauged under the electroweak group. Although near the conformal window 
\cite{Appelquist:1998xf,Sundrum:1991rf} the $S$ parameter is reduced due to
non-perturbative corrections, it is still too large if the
model has a large particle content. In addition, such models may
have a large number of pseudo Nambu-Goldstone bosons. By
choosing a higher dimensional technicolor representation for the new technifermions one
can overcome these problems \cite{Sannino:2004qp,Hong:2004td}. 

To have a very low $S$ parameter one would ideally have a technicolor theory which with only one doublet breaks dynamically the electroweak symmetry but at the same time being walking (near conformal) to reduce the $S$ parameter. The walking nature then also enhances the scale responsible for the fermion mass generation. 
 
According to the phase diagram exhibited earlier the promising candidate theories with the properties required are either theories with fermions in the adjoint representation or two index symmetric one.

The relevant feature, found first in \cite{Sannino:2004qp} is that the
S-type theories can be near conformal already at $N_{T f}=2$ when $N=2$ or $3$. This
should be contrasted with theories in which the fermions are in the fundamental
representation for which the minimum number of flavors required to
reach the conformal window is eight for $N=2$. The critical value of flavors increases with the number of colors
for the gauge theory with S-type matter: the limiting value is
$4.15$ at large $N$. 

We refer with minimal theories for which the number of flavors needed to achieve an infrared fixed point is very small compared to the case of matter in the fundamental representation of the gauge group. 

\subsection{Minimal Walking Technicolor (MWT)}

The dynamical sector we consider, which underlies the Higgs mechanism, is an SU(2) technicolor gauge theory with two adjoint
technifermions \cite{Sannino:2004qp}. The theory is asymptotically free if the number of flavors $N_f$ is less than $2.75$ according to the ladder approximation. The two adjoint fermions are conveniently written as \beq Q_L^a=\left(\begin{array}{c} U^{a} \\D^{a} \end{array}\right)_L , \qquad U_R^a \
, \quad D_R^a \ ,  \qquad a=1,2,3 \ ,\eeq with $a$ being the adjoint color index of SU(2). The left handed fields are arranged in three
doublets of the SU(2)$_L$ weak interactions in the standard fashion. The condensate is $\langle \bar{U}U + \bar{D}D \rangle$ which correctly breaks the electroweak symmetry.

The model as described so far suffers from the Witten topological anomaly. This can be fixed by
adding a new weakly charged fermionic doublet which is a technicolor singlet \cite{Dietrich:2005jn}. Schematically: 
\beq L_L =
\left(
\begin{array}{c} N \\ E \end{array} \right)_L , \qquad N_R \ ,~E_R \
. \eeq 
The low-energy effective theory to be tested at the LHC, the comparison with precision data and a first study of the unitarity of $WW$ longitudinal scattering can be found in \cite{Foadi:2007ue,Foadi:2007se,Foadi:2008ci}. In \cite{Gudnason:2006mk} we discussed the unification issue within this model. {}Further studies appeared in \cite{Chen:2008jn, Christensen:2005bt}. 
\subsection{Next to Minimal Walking Technicolor Theory}
\label{4}

The theory with three technicolors contains an even number of electroweak doublets, and hence
it is not subject to a Witten anomaly.  
The doublet of technifermions, is then represented again as:
\bear
Q_L^{\{C_1,C_2 \}}
=
\left(\begin{array}{l}U^{\{C_1,C_2 \}}\\ D^{\{C_1,C_2 \}}\end{array}\right)_L \ ,
\qquad 
Q_R^{\{C_1,C_2\}}&=&\left(U_R^{\{C_1,C_2\}},~ D_R^{\{C_1,C_2\}}\right) \ . \nonumber 
\eear
Here $C_i=1,2,3$ is the technicolor index and $Q_{L(R)}$ is a doublet (singlet) with respect 
to the weak interactions. 
Since the two-index symmetric representation of $SU(3)$ is complex the flavor symmetry is $SU(2)_L\times SU(2)_R\times U(1)$. 
Only three Goldstones emerge and
are absorbed in the longitudinal components of the weak vector bosons. {}More information about this theory can be found in \cite{Dietrich:2005wk}.

\subsection{Partially Gauged Technicolor (PGT)}

A small modification of the traditional technicolor approach, which neither
involves additional particle species nor more complicated gauge groups, 
allows constructing several other viable candidates. It consists in letting 
only one doublet of techniquarks transform non-trivially under the electroweak
symmetries with the rest being electroweak singlets, as first suggested in 
\cite{Dietrich:2005jn} and later also used in \cite{Christensen:2005cb}.

Still, all techniquarks transform under the technicolor gauge group. Thereby only one techniquark doublet contributes directly to the oblique 
parameter which is thus kept to a minimum for theories which need more
than one family of techniquarks to be quasi-conformal. It is the condensation 
of that first electroweakly charged family that breaks the electroweak 
symmetry. We provided in \cite{Dietrich:2006cm} an exhaustive list, given the knowledge about the phase diagram, of the possible underlying gauge theories one can use to construct PGT models. It is obvious that to be phenomenologically viable PGT requires the introduction, by hand, of mass terms for the flavors not gauged under the electroweak symmetry. The simplest model is an $SU(N)$  gauge theory with a number of flavors in the fundamental representations sufficiently large that the massless theory is (near) conformal. 

\subsection{Conformal Technicolor = PGT}

 Luty  in \cite{Luty:2008vs} constructed a model of conformal technicolor \cite{Luty:2004ye} using, in practice, the PGT model described above. We repeat once more that the addition of a mass term for the flavors not gauged under the electroweak symmetry  is a necessity for any phenomenologically viable PGT model. In fact if the underlying PGT is near conformal the large chiral symmetry group breaks spontaneously and one must give mass to the phenomenologically unacceptable electroweak neutral Goldstone bosons. If the underlying theory is conformal a mass term must be introduced as well to generate the scale responsible for the breaking the electroweak symmetry in the first place. In \cite{Dietrich:2006cm} we discussed the precision constraints for PGT  while the bound of the large anomalous dimensions for the fermion condensate and its impact on the conformal window for nonsupersymmetric theories as well as the generation of a realistic top mass is present below equation (17) of \cite{Ryttov:2007cx}.

\section{Conclusions}
I reviewed new ideas, tools and results enabling us to give vital information for plotting the phase diagram for strongly coupled theories. I have also presented, with the help of the phase diagram, prime candidates for dynamical models of electroweak symmetry breaking. 

\section{Acknowledgments}

 I am partially supported by the Marie Curie Excellence Grant under contract MEXT-CT-2004-013510.


\begin{thebibliography}{99}

\bibitem{Weinberg:1979bn}
  S.~Weinberg,
  Phys.\ Rev.\  D {\bf 19}, 1277 (1979).

\bibitem{Susskind:1978ms}
  L.~Susskind,
  Phys.\ Rev.\  D {\bf 20}, 2619 (1979).

\bibitem{Sannino:2008ha}
  F.~Sannino,
  arXiv:0804.0182 [hep-ph].

\bibitem{Hill:2002ap}
  C.~T.~Hill and E.~H.~Simmons,
  Phys.\ Rept.\  {\bf 381}, 235 (2003)
  [Erratum-ibid.\  {\bf 390}, 553 (2004)]
  [arXiv:hep-ph/0203079].

\bibitem{Lane:2002wv}
  K.~Lane,
  arXiv:hep-ph/0202255.

\bibitem{Ryttov:2007cx}
  T.~A.~Ryttov and F.~Sannino,
  arXiv:0711.3745 [hep-th].

\bibitem{Ryttov:2007sr}
  T.~A.~Ryttov and F.~Sannino,
  Phys.\ Rev.\  D {\bf 76}, 105004 (2007)
  [arXiv:0707.3166 [hep-th]].

\bibitem{Dietrich:2006cm}
  D.~D.~Dietrich and F.~Sannino,
  Phys.\ Rev.\  D {\bf 75}, 085018 (2007)
  [arXiv:hep-ph/0611341].

\bibitem{Sannino:2004qp}
  F.~Sannino and K.~Tuominen,
  Phys.\ Rev.\  D {\bf 71}, 051901 (2005)
  [arXiv:hep-ph/0405209].

\bibitem{Dietrich:2005jn}
  D.~D.~Dietrich, F.~Sannino and K.~Tuominen,
  Phys.\ Rev.\  D {\bf 72}, 055001 (2005)
  [arXiv:hep-ph/0505059].

\bibitem{Dietrich:2005wk}
  D.~D.~Dietrich, F.~Sannino and K.~Tuominen,
  Phys.\ Rev.\  D {\bf 73}, 037701 (2006)
  [arXiv:hep-ph/0510217].

\bibitem{Foadi:2007ue}
  R.~Foadi, M.~T.~Frandsen, T.~A.~Ryttov and F.~Sannino,
  Phys.\ Rev.\  D {\bf 76}, 055005 (2007)
  [arXiv:0706.1696 [hep-ph]].

\bibitem{DelDebbio:2008wb}
  L.~Del Debbio, M.~T.~Frandsen, H.~Panagopoulos and F.~Sannino,
  JHEP {\bf 0806}, 007 (2008)
  [arXiv:0802.0891 [hep-lat]].

\bibitem{Catterall:2007yx}
  S.~Catterall and F.~Sannino,
  Phys.\ Rev.\  D {\bf 76}, 034504 (2007)
  [arXiv:0705.1664 [hep-lat]].

\bibitem{Appelquist:2007hu}
  T.~Appelquist, G.~T.~Fleming and E.~T.~Neil,
  Phys.\ Rev.\ Lett.\  {\bf 100}, 171607 (2008)
  [arXiv:0712.0609 [hep-ph]].

\bibitem{Appelquist:1988yc}
  T.~Appelquist, K.~D.~Lane and U.~Mahanta,
  Phys.\ Rev.\ Lett.\  {\bf 61}, 1553 (1988).

\bibitem{Cohen:1988sq}
  A.~G.~Cohen and H.~Georgi,
  Nucl.\ Phys.\  B {\bf 314}, 7 (1989).

\bibitem{Miransky:1996pd}
  V.~A.~Miransky and K.~Yamawaki,
  Phys.\ Rev.\  D {\bf 55}, 5051 (1997)
  [Erratum-ibid.\  D {\bf 56}, 3768 (1997)]
  [arXiv:hep-th/9611142].

\bibitem{Appelquist:1999hr}
  T.~Appelquist, A.~G.~Cohen and M.~Schmaltz,
  Phys.\ Rev.\  D {\bf 60}, 045003 (1999)
  [arXiv:hep-th/9901109].

\bibitem{Sannino:2005sk}
  F.~Sannino,
  Phys.\ Rev.\  D {\bf 72}, 125006 (2005)
  [arXiv:hep-th/0507251].

\bibitem{Appelquist:1999vs}
  T.~Appelquist, A.~G.~Cohen, M.~Schmaltz and R.~Shrock,
  Phys.\ Lett.\  B {\bf 459}, 235 (1999)
  [arXiv:hep-th/9904172].

\bibitem{Appelquist:2000qg}
  T.~Appelquist, Z.~y.~Duan and F.~Sannino,
  Phys.\ Rev.\  D {\bf 61}, 125009 (2000)
  [arXiv:hep-ph/0001043].

\bibitem{Novikov:1983uc}
  V.~A.~Novikov, M.~A.~Shifman, A.~I.~Vainshtein and V.~I.~Zakharov,
  Nucl.\ Phys.\  B {\bf 229}, 381 (1983).

\bibitem{Shifman:1986zi}
  M.~A.~Shifman and A.~I.~Vainshtein,
  Nucl.\ Phys.\  B {\bf 277}, 456 (1986)
  [Sov.\ Phys.\ JETP {\bf 64}, 428 (1986\ ZETFA,91,723-744.1986)].

\bibitem{Banks:1981nn}
  T.~Banks and A.~Zaks,
  Nucl.\ Phys.\  B {\bf 196}, 189 (1982).

\bibitem{Mack:1975je}
  G.~Mack,
  Commun.\ Math.\ Phys.\  {\bf 55}, 1 (1977).

\bibitem{Flato:1983te}
  M.~Flato and C.~Fronsdal,
  Lett.\ Math.\ Phys.\  {\bf 8}, 159 (1984).

\bibitem{Dobrev:1985qv}
  V.~K.~Dobrev and V.~B.~Petkova,
  Phys.\ Lett.\  B {\bf 162}, 127 (1985).

\bibitem{Seiberg:1994pq}
  N.~Seiberg,
  Nucl.\ Phys.\  B {\bf 435}, 129 (1995)
  [arXiv:hep-th/9411149].

\bibitem{Yamawaki:1985zg}
  K.~Yamawaki, M.~Bando and K.~i.~Matumoto,
  Phys.\ Rev.\ Lett.\  {\bf 56}, 1335 (1986).

\bibitem{Holdom:1984sk}
  B.~Holdom,
  Phys.\ Lett.\  B {\bf 150}, 301 (1985).

\bibitem{Holdom:1981rm}
  B.~Holdom,
  Phys.\ Rev.\  D {\bf 24}, 1441 (1981).

\bibitem{Appelquist:1986an}
  T.~W.~Appelquist, D.~Karabali and L.~C.~R.~Wijewardhana,
  Phys.\ Rev.\ Lett.\  {\bf 57}, 957 (1986).

\bibitem{Evans:2005pu}
  N.~Evans and F.~Sannino,
  arXiv:hep-ph/0512080.

\bibitem{Shamir:2008pb}
  Y.~Shamir, B.~Svetitsky and T.~DeGrand,
  arXiv:0803.1707 [hep-lat].

\bibitem{DelDebbio:2008zf}
  L.~Del Debbio, A.~Patella and C.~Pica,
  arXiv:0805.2058 [hep-lat].

\bibitem{Deuzeman:2008sc}
  A.~Deuzeman, M.~P.~Lombardo and E.~Pallante,
  arXiv:0804.2905 [hep-lat].

\bibitem{Iwasaki:2003de}
  Y.~Iwasaki, K.~Kanaya, S.~Kaya, S.~Sakai and T.~Yoshie,
  Phys.\ Rev.\  D {\bf 69}, 014507 (2004)
  [arXiv:hep-lat/0309159].

\bibitem{Luscher:1992an}
  M.~Luscher, R.~Narayanan, P.~Weisz and U.~Wolff,
  Nucl.\ Phys.\  B {\bf 384}, 168 (1992)
  [arXiv:hep-lat/9207009].

\bibitem{Luscher:1993gh}
  M.~Luscher, R.~Sommer, P.~Weisz and U.~Wolff,
  Nucl.\ Phys.\  B {\bf 413}, 481 (1994)
  [arXiv:hep-lat/9309005].

\bibitem{Sint:1993un}
  S.~Sint,
  Nucl.\ Phys.\  B {\bf 421}, 135 (1994)
  [arXiv:hep-lat/9312079].

\bibitem{Bode:1999sm}
  A.~Bode, P.~Weisz and U.~Wolff  [ALPHA collaboration],
  Nucl.\ Phys.\  B {\bf 576}, 517 (2000)
  [Erratum-ibid.\  B {\bf 600}, 453 (2001\ ERRAT,B608,481.2001)]
  [arXiv:hep-lat/9911018].

\bibitem{Sannino:1999qe}
  F.~Sannino and J.~Schechter,
  Phys.\ Rev.\  D {\bf 60}, 056004 (1999)
  [arXiv:hep-ph/9903359].

\bibitem{Luty:2008vs}
  M.~A.~Luty,
  arXiv:0806.1235 [hep-ph].

\bibitem{Hong:2004td}
  D.~K.~Hong, S.~D.~H.~Hsu and F.~Sannino,
  Phys.\ Lett.\  B {\bf 597}, 89 (2004)
  [arXiv:hep-ph/0406200].

\bibitem{Appelquist:1998xf}
  T.~Appelquist and F.~Sannino,
  Phys.\ Rev.\  D {\bf 59}, 067702 (1999)
  [arXiv:hep-ph/9806409].

\bibitem{Sundrum:1991rf}
  R.~Sundrum and S.~D.~H.~Hsu,
  Nucl.\ Phys.\  B {\bf 391}, 127 (1993)
  [arXiv:hep-ph/9206225].

\bibitem{Foadi:2007se}
  R.~Foadi, M.~T.~Frandsen and F.~Sannino,
  Phys.\ Rev.\  D {\bf 77}, 097702 (2008)
  [arXiv:0712.1948 [hep-ph]].

\bibitem{Foadi:2008ci}
  R.~Foadi and F.~Sannino,
  arXiv:0801.0663 [hep-ph].

\bibitem{Gudnason:2006mk}
  S.~B.~Gudnason, T.~A.~Ryttov and F.~Sannino,
  Phys.\ Rev.\  D {\bf 76}, 015005 (2007)
  [arXiv:hep-ph/0612230].

\bibitem{Chen:2008jn}
  N.~Chen and R.~Shrock,
  arXiv:0805.3687 [hep-ph].

\bibitem{Christensen:2005bt}
  N.~D.~Christensen and R.~Shrock,
  Phys.\ Rev.\  D {\bf 72}, 035013 (2005)
  [arXiv:hep-ph/0506155].

\bibitem{Christensen:2005cb}
  N.~D.~Christensen and R.~Shrock,
  Phys.\ Lett.\  B {\bf 632}, 92 (2006)
  [arXiv:hep-ph/0509109].

\bibitem{Luty:2004ye}
  M.~A.~Luty and T.~Okui,
  JHEP {\bf 0609}, 070 (2006)
  [arXiv:hep-ph/0409274].
\end{thebibliography}
\end{document}